\documentclass[conference]{IEEEtran}

\usepackage{graphicx}
\usepackage{amsmath,amssymb,amsfonts}
\usepackage{multirow}
\usepackage{psfrag}
\usepackage{subfigure}
\usepackage{caption}
\usepackage{wrapfig}
\usepackage{color}
\usepackage{soul}
\usepackage{xcolor}
\usepackage[bookmarks=false]{hyperref}
\hypersetup{breaklinks=true}
\usepackage{lipsum}
\usepackage[ruled,vlined,linesnumbered]{algorithm2e}
\usepackage{mathrsfs}
\usepackage{pbox}

\DontPrintSemicolon

\newcommand{\matr}[1]{\mathbf{#1}}

\makeatletter
\def\hlinewd#1{%
  \noalign{\ifnum0=`}\fi\hrule \@height #1 \futurelet
   \reserved@a\@xhline}
\makeatother

\graphicspath{{Figures/}}
\ifCLASSINFOpdf
\else
\fi

\begin{document}

\title{DeepWiTraffic: Low Cost WiFi-Based Traffic Monitoring System Using Deep Learning}

\author{Myounggyu~Won$^{1}$, Sayan~Sahu$^{2}$, and Kyung-Joon Park$^3$\\
$^1$CS$^3$ Lab, University of Memphis, TN, United States\\
$^2$EECS Department, South Dakota State University, SD, United States\\
$^3$CPS Global Center, Daegu Gyeongbuk Institute of Science and Technology, Daegu, South Korea\\
mwon@memphis.edu, sayan.sahu@sdstate.edu, kjp@dgist.ac.kr
}

\markboth{Journal of \LaTeX\ Class Files,~Vol.~13, No.~9, September~2014}%
{Shell \MakeLowercase{\textit{et al.}}: Bare Demo of IEEEtran.cls for Journals}

\maketitle

\begin{abstract}
A traffic monitoring system (TMS) is an integral part of Intelligent Transportation Systems (ITS). It is an essential tool for traffic analysis and planning. One of the biggest challenges is, however, the high cost especially in covering the huge rural road network. In this paper, we propose to address the problem by developing a novel TMS called $\mathsf{DeepWiTraffic}$. $\mathsf{DeepWiTraffic}$ is a low-cost, portable, and non-intrusive solution that is built only with two WiFi transceivers. It exploits the unique WiFi Channel State Information (CSI) of passing vehicles to perform detection and classification of vehicles. Spatial and temporal correlations of CSI amplitude and phase data are identified and analyzed using a machine learning technique to classify vehicles into five different types: motorcycles, passenger vehicles, SUVs, pickup trucks, and large trucks. A large amount of CSI data and ground-truth video data are collected over a month period from a real-world two-lane rural roadway to validate the effectiveness of $\mathsf{DeepWiTraffic}$. The results validate that $\mathsf{DeepWiTraffic}$ is an effective TMS with the average detection accuracy of 99.4\% and the average classification accuracy of 91.1\% in comparison with state-of-the-art non-intrusive TMSs.

\end{abstract}


\IEEEpeerreviewmaketitle

\section{Introduction}
\label{sec:introduction}

A traffic monitoring system (TMS) is an important component of Intelligent Transportation Systems (ITS). It is deployed on public roads to collect traffic data to characterize the performance of a roadway system. Providing quality, timely, and complete traffic volume and classification data is a key functionality of a TMS~\cite{gdot}. These traffic data are essential in analyzing transportation systems for more effective utilization of resources, improving environmental sustainability, and estimating future transportation needs including road improvement, assessment of road network efficiency, and analysis of economic benefits,~\emph{etc.} ~\cite{mimbela2007summary}.

For example, in the U.S., the Department of Transportation (DOT) of each state is charged by the United States Federal Highway
Administration (FHWA) to collect traffic information about vehicles traveling state and federal highways and roadways to improve the safety and efficiency of transportation~\cite{trafficManagement}. As such, state highway and transportation agencies operate TMSs to perform vehicle counting, vehicle classification, and vehicle weight measurement. These TMSs are either temporary or permanent. There are 7,430 TMSs under operation in the U.S. as of August 2015~\cite{trafficMonitoringLocations}.

A critical problem of deploying a TMS is, however, the high cost to cover the huge rural road network. For example, in the U.S., there are over 100,000 miles of rural highways. Yet, the estimated cost is non-negligible. According to the Georgia DOT, the minimum cost to install a continuous TMS on a two-lane rural roadway is about \$25,000~\cite{challenge}, and 365 day vehicle classification on a two-lane rural roadway is more expensive costing about \$35,770~\cite{trafficMonitoringGuide}. \emph{This paper specifically aims at addressing this endemic cost issue by developing an innovative TMS targeting typical two-lane rural roadways.}

Vehicle detection and classification techniques are largely categorized into three types depending on where sensors are installed: intrusive, non-intrusive, and off-roadway~\cite{balid2018intelligent}. Intrusive solutions embed sensors such as magnetic detectors~\cite{xu2018vehicle}, vibration sensors~\cite{stocker2014situational}, and inductive loops~\cite{jeng2014high} in the pavement of roadway. Intrusive solutions, however, incur very high cost for installation and maintenance because sensors are installed under pavement surface which requires traffic disruption and lane closure. As such, non-intrusive solutions have been increasingly adopted. Non-intrusive approaches mount sensors like magnetic sensors~\cite{yang2015vehicle}, acoustic sensors~\cite{george2013vehicle}, and LIDAR (Laser Infrared Detection And Ranging)~\cite{lee2015using} either on roadsides or over the road. One of the most widely used sensors in this type of solution is a camera. The performance of a camera-based solution is, however, degraded when vision obstructions are present, and even severely under adverse weather conditions. Furthermore, camera-based solutions raise the privacy issue. Other sensors for non-intrusive solutions such as magenetic sensors and acoustic sensors require precise calibration of sensor direction and placement~\cite{odat2018vehicle}. Off-roadway solutions use mobile sensor systems such as UAVs~\cite{liu2015fast}\cite{tang2017arbitrary} or satellites~\cite{audebert2017segment}. Especially due to recent advances in UAV technologies, off-road-based approaches are receiving greater attention. However, off-roadway-based solutions suffer from spatial and temporal limitations, \emph{e.g.,} the operation time of a UAV is constrained by the limited flight time, and satellites are not always available.


In this paper, we develop a low-cost, portable, and non-intrusive TMS with an aim to overcome the limitations of existing approaches specifically concentrating on the cost issue. The proposed TMS hinges on distinctive wireless channel characteristics induced by passing vehicles to classify them into five vehicle types: motorcycles, passenger cars, SUVs, pickup trucks, and large trucks. $\mathsf{DeepWiTraffic}$ utilizes WiFi channel state information (CSI) that conveys rich information about the changes in the channel properties caused by passing vehicles. Especially the spatial and temporal correlations of CSI phase and amplitude of different subcarriers are analyzed for effective vehicle classification. A convolutional neural network (CNN) is designed to capture the optimal features of CSI data automatically and train the vehicle classification model.

Numerous technical challenges are addressed to achieve high vehicle detection/classification accuracy at a significantly reduced cost. The environment noise in CSI amplitude data caused by surrounding obstacles and low-speed moving objects, \emph{e.g.,} people moving around are effectively mitigated. Principal component analysis (PCA) is exploited to reduce the dimension of multiple subcarriers (in our experiments, 30 subcarriers for each TX and RX pair) down to one, expediting the processing speed for classification to cope with fast vehicle speed and sharpening the vehicle detection performance. A linear transformation-based phase preprocessing technique is designed to effectively capture the changes in CSI phase data induced by passing vehicles. Consequently, an effective convolutional neural network (CNN) is designed. The preprocessed CSI data are formatted into an image and is provided as input to CNN to generate a vehicle classification model and to perform vehicle classification.

We collected a large amount of CSI data and ground-truth video data over a month period. Rigorous experiments were performed with various combinations of hyper parameters in training the CNN model to improve the classification accuracy. The results show that the average vehicle detection and classification accuracy of $\mathsf{DeepWiTraffic}$ were 99.4\% and 91.1\%, respectively.



\section{Preliminaries and Problem Statement}
\label{sec:preliminaries}

\subsection{WiFi Channel State Information}
\label{subsec:wifi_channel_state_information}


The orthogonal frequency division multiplexting (OFDM) modulation scheme is used to implement the physical layer of contemporary WiFi standards~\cite{hanzo2010mimo}. It is robust against frequency selective fading since a high data-rate stream is partitioned onto close-spaced subcarriers. WiFi CSI represents the channel properties for the OFDM subcarriers, \emph{i.e.,} a combined effect of fading, scattering, and power decay with distance. Formally, the channel properties are modeled as follows: $y=H \cdot x + n$~\cite{zeng2014your}, where $x$ and $y$ refer to the transmitted and received signals, respectively; $n$ is the channel noise, and $H$ is a $h_{tx} \times h_{rx} \times t_{sub}$ matrix, where $h_{tx}$, $h_{rx}$, and $h_{sub}$, are the number of receiver antennas, transmitter antennas, and subcarriers, respectively. The matrix $H$ is the WiFi CSI and can be expressed as a vector of $h_{sub}$ subcarrier groups as follows: $H=[H_1, H_2, ..., H_{h_{sub}}]$, where $H_i$ is a $h_{tx} \times h_{rx}$ matrix which contains the WiFi CSI value for the $i$-th subcarrier received via $h_{tx} \times h_{rx}$ different transmitter-receiver antenna pairs. A CSI value for the $i$-th subcarrier received via a TX antenna $k$ and receiver antenna $l$ pair is denoted by $CSI^{i}$, which is defined as follows: $CSI^{i}_{kl} = |\eta|e^{j\phi}$, where $|\eta|$ is the amplitude and $\phi$ is the phase information of the CSI value.

\subsection{Problem Statement}
\label{sub:problem_statement}

Let $M_{CSI}$ denote a $N \times 30 \times N_{A}$ matrix where each element represents a CSI value. Here $N$ is the number of successively received packets; $N_{A}$ is the number of TX-RX antenna pairs; and the number 30 means the number of subcarriers (In our experiment, 30 subcarriers are used). Thus, $M_{CSI}$ is a data structure that maintains all CSI values for $N$ successively received packets where each packet is transmitted through $30 \times N_{A}$ subcarriers. Also, denote $N_C$ be the total number of vehicles passed. We are tasked to classify $N_C$ vehicles into five different types \{bike, passenger car, SUV, pickup truck, large truck\} given $M_{CSI}$ as input.

The CSI amplitude and phase values are extracted from $M_{CSI}$. Extracted amplitude and phase values are denoted by sets $A=\{a_1,...,a_{N}\}$, and $P=\{p_1,...,p_N\}$, respectively. A vehicle detection algorithm $\mathscr{A}_d$ detects $i$-th vehicle and extracts from $A$ (and $P$) a set of ``induced'' CSI amplitude (and phase) values corresponding to the detected $i$-th vehicle, which are denoted by $\hat{A}_i$ (and $\hat{P}_i$). Now we can create a collection of $\hat{A}_i$ and $\hat{P}_i$ for all passing vehicles $i$, $0 \le i \le N_C$, which is denoted by $\matr{A}=\{\hat{A}_1, ..., \hat{A}_{N_C}\}$ and $\matr{P}=\{\hat{P}_1, ..., \hat{P}_{N_C}\}$, respectively. These collections $\matr{A}$ and $\matr{P}$ are provided as input to a convolutional neural network (CNN) to train a model $\mathscr{M}$ (in training mode) and to classify based on input CSI data consisting of $\hat{A}_i$ and $\hat{P}_i$ into five vehicle types \{bike, passenger car, SUV, pickup truck, large truck\} using the model (in testing mode). The algorithm used for this vehicle classification is denoted by $\mathscr{A}_c$.

In this paper, we develop the two algorithms namely $\mathscr{A}_d$ and $\mathscr{A}_c$ especially targeting passing vehicles on typical two-lane rural roadways. In subsequent sections, we will describe (1) CSI data preprocessing techniques that are designed to reduce the noise and dimension of raw CSI amplitude and phase data for faster and more effective processing, (2) algorithms to extract the ``induced'' CSI amplitude and phase values corresponding to a passing vehicle, and (3) design of a deep learning network for effective vehicle classification. Notations used throughout this paper are listed in Table II.

\section{Proposed System}
\label{sec:proposed_system}

\subsection{System Overview}
\label{sec:system_overview}

\begin{figure}[!htbp]
\centering
\includegraphics[width=.8\columnwidth]{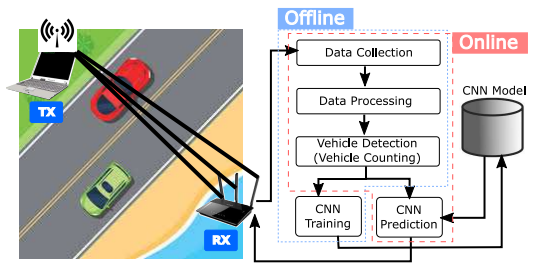}
\caption {System architecture of $\mathsf{DeepWiTraffic}$.}
\label{fig:system_overview}
\end{figure}

$\mathsf{DeepWiTraffic}$ consists of four system components, namely Data Collection, Data Processing, Vehicle Detection, and Vehicle Classification. Figure~\ref{fig:system_overview} displays the system architecture of $\mathsf{DeepWiTraffic}$. The Data Collection module constructs $M_{CSI}$ with $N$ successively received packets. The Data Processing module handles three tasks: extraction of CSI amplitude $A$ and phase $P$ from $M_{CSI}$, noise removal from $A$ and $P$, dimension reduction of $A$ based on principal component analysis for faster classification. The Vehicle Detection module implements $\mathscr{A}_d$. It detects a passing vehicle and extracts the corresponding CSI amplitude $\matr{A}$ and phase values $\matr{P}$ from $A$ and $P$, respectively. The Vehicle Classification module performs two tasks: CNN training and CNN prediction. In the former task, a CNN model $\mathscr{M}$ is created using $\matr{A}$ and $\matr{P}$ as input. In the latter task, the module performs vehicle classification.

\subsection{CSI Data Processing}
\label{sec:noise_removal}

\subsubsection{Low Pass Filtering}
\label{sec:low_pass_filter}

\begin{figure}[!htbp]
\begin{minipage}[b]{0.48\columnwidth}
\centering
\includegraphics[width=\columnwidth]{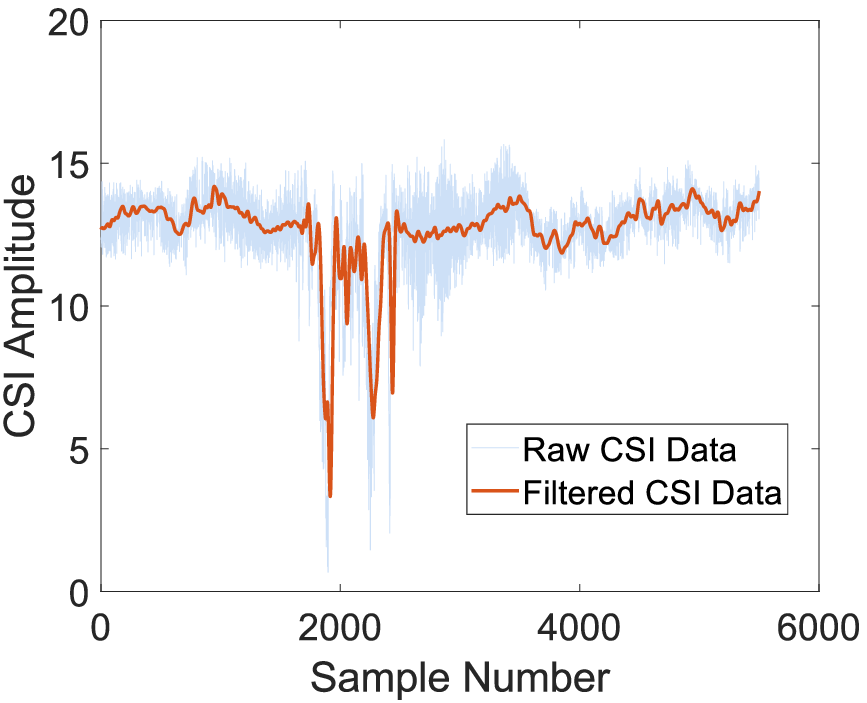}
\caption {Raw vs filtered CSI data of a passing vehicle.}
\label{fig:filtered}
\end{minipage}
\hspace{1mm}
\begin{minipage}[b]{0.48\columnwidth}
\centering
\includegraphics[width=\columnwidth]{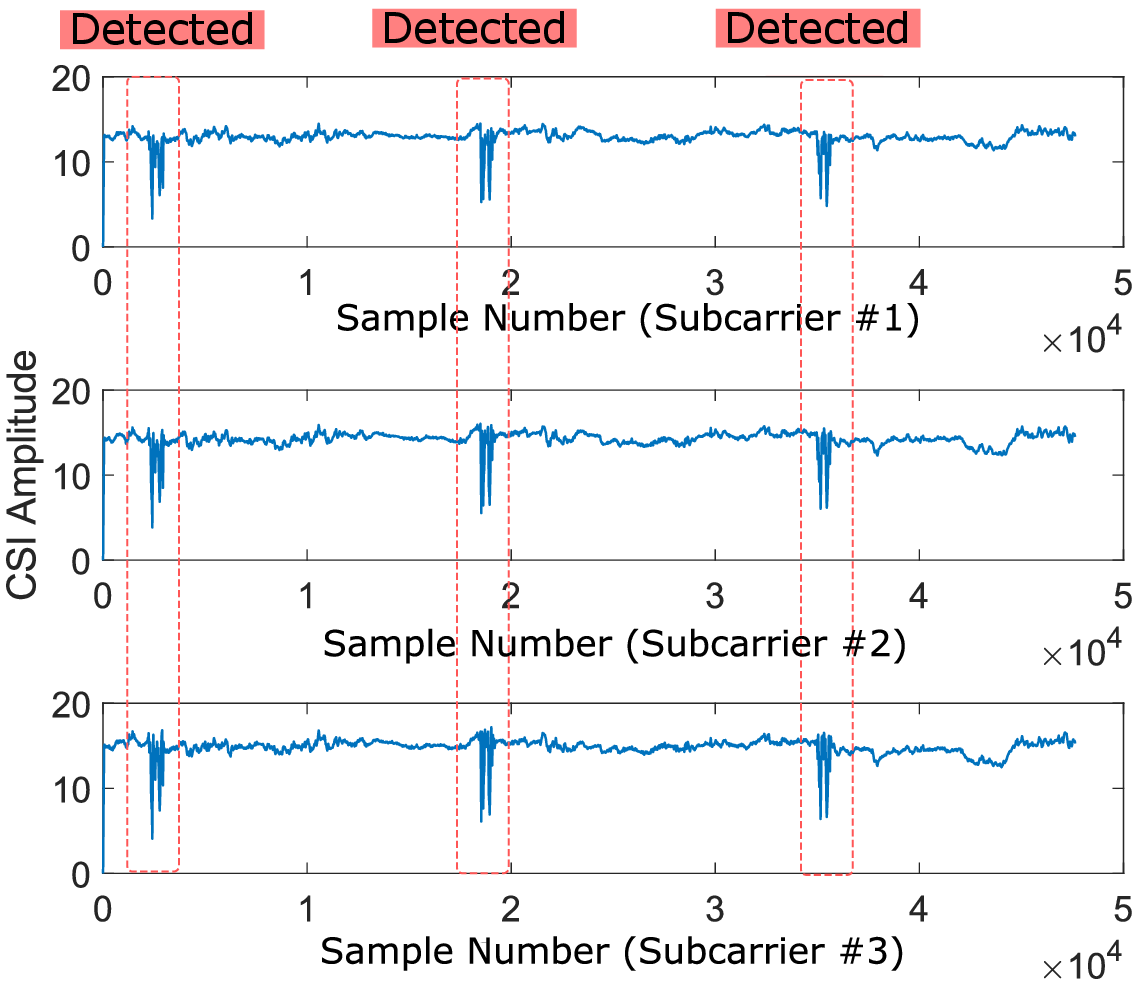}
\caption {Filtered WiFi CSI streams for subcarriers.}
\label{fig:correlated}
\end{minipage}
\end{figure}


Collected CSI data contain noise which disturbs effective vehicle detection and classification. In addition to reducing the static background noise, we focus on reducing noise caused by slow-moving objects around the system such as a person, in most cases the motion of the system operators. More precisely, we mitigate the noise that is induced by any object that moves at a speed of less than 2m/s to capture exclusively the CSI data corresponding to fast-moving cars. The WiFi wavelength of our system that operates at 5.32GHz frequency bandwidth is 5.64cm~\cite{wang2015understanding}. With the wavelength of 5.64cm and the speed of 2m/s, the corresponding frequency component is calculated as 38Hz. As such, we apply a standard low pass filter with a cutoff frequency of 38Hz. Figure~\ref{fig:filtered} shows an example of raw CSI data and filtered counterpart. As it is shown, the noise has been effectively mitigated.

\subsubsection{PCA-Based Denoising}
\label{sec:pca}

Environmental noise (\emph{e.g.,} caused by slow moving objects) has been successfully mitigated by applying a low pass filter. Another important source of performance degradation is the noise caused by internal state transitions in a WiFi NIC which include changes in transmission power, adaptation of transmission rate, and CSI reference level changes~\cite{li2016csi}. Typically, burst noises in CSI data are caused by these internal state transitions. Ali \emph{et al.} made an interesting observation that the effect of these burst noises is significantly correlated across CSI data streams of subcarriers~\cite{ali2015keystroke}.


The principal component analysis (PCA) is used to mitigate the burst noises by exploiting highly correlated CSI streams for different subcarriers. Figure~\ref{fig:correlated} depicts an example illustrating that CSI streams for different subcarriers are highly correlated. The PCA is also used to reduce the dimension of CSI data from 30 streams down to one for faster processing. This dimension reduction is especially important for expediting the vehicle classification process since cars pass the system very quickly.  More specifically, using PCA, we analyze the correlations of these multi-dimensional CSI data, extract common features, and reduce the dimension to one. This noise and dimension reduction process is executed in four steps as follows.

\textbf{Preprocessing of Sample:} For each TX-RX antenna pair, define a $N \times 30$ matrix denoted by $H_{CSI,AMP}$ that stores CSI amplitude values for $N$ successive packets for 30 subcarriers. A CSI stream (consisting of $N$ CSI amplitude values) for each subcarrier is arranged in each column of matrix $H_{CSI,AMP}$. After construction of matrix $H_{CSI,AMP}$, the mean value of each column is calculated and subtracted from each column, which completes this step.

\textbf{Computation of Covariance Matrix:} the covariance matrix $H_{CSI,AMP}^{T} \times H_{CSI,AMP}$ is calculated in this step.

\textbf{Computation of Eigenvalues and Eigenvectors of Covariance:} Eigendecomposition of the covariance matrix $H_{CSI,AMP}^{T} \times H_{CSI,AMP}$ is performed to obtain the eigenvectors $q$ ($30 \times k$).

\textbf{Reconstruction of Signal:} By projecting $H_{CSI,AMP}$ onto the eigenvectors $q$ ($30 \times k$), we obtain $h_i = H_{CSI,AMP} \times q_i$, where $q_i$ is the $i^{th}$ eigenvector and $h_i$ is the $i^{th}$ principal component.

\begin{figure}[!htbp]
\begin{minipage}[b]{0.48\columnwidth}
\centering
\includegraphics[width=\columnwidth]{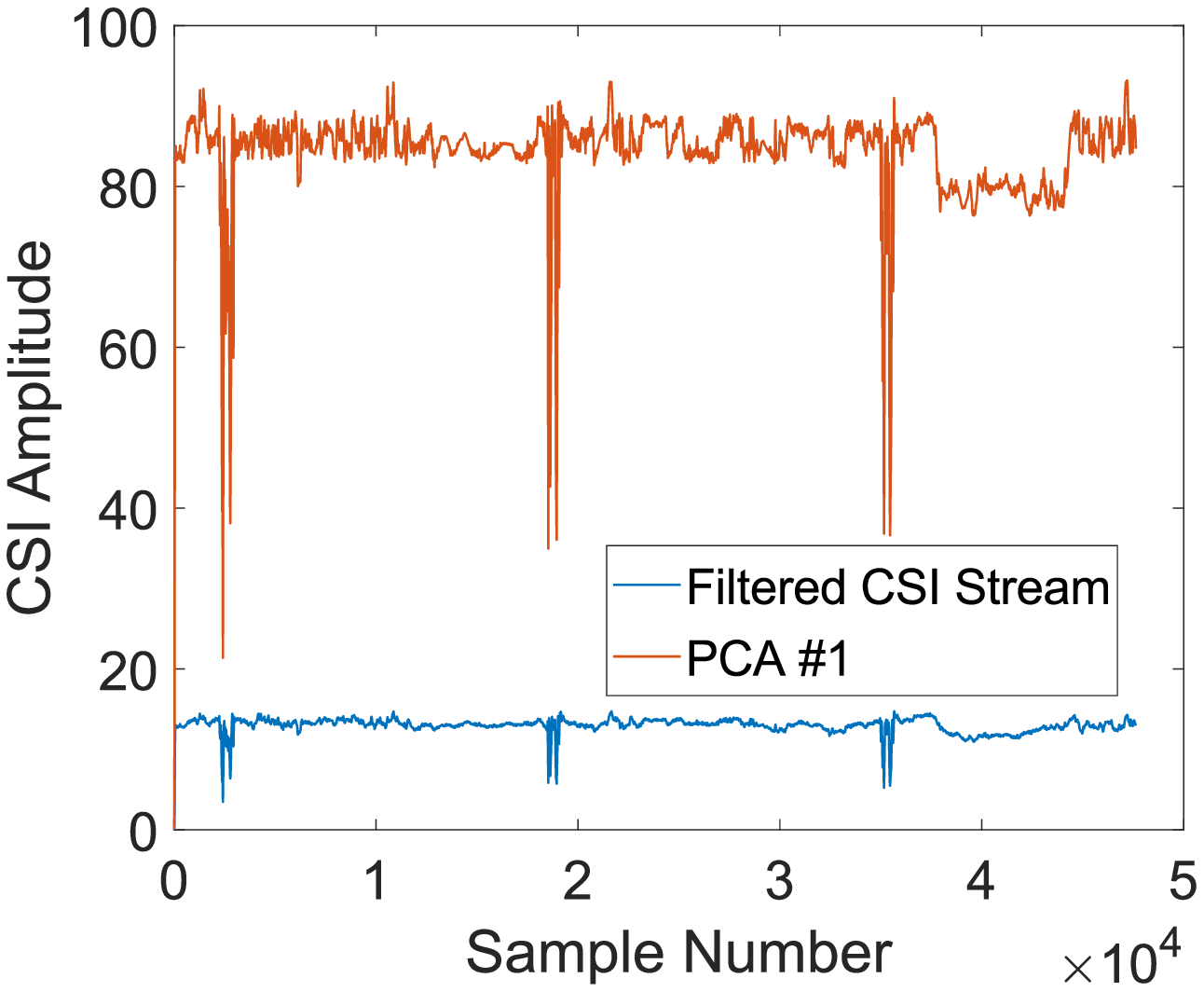}
\caption {PCA \#1 that represents all 30 CSI streams vs a CSI stream.}
\label{fig:pca_comp}
\end{minipage}
\hspace{1mm}
\begin{minipage}[b]{0.48\columnwidth}
\centering
\includegraphics[width=\columnwidth]{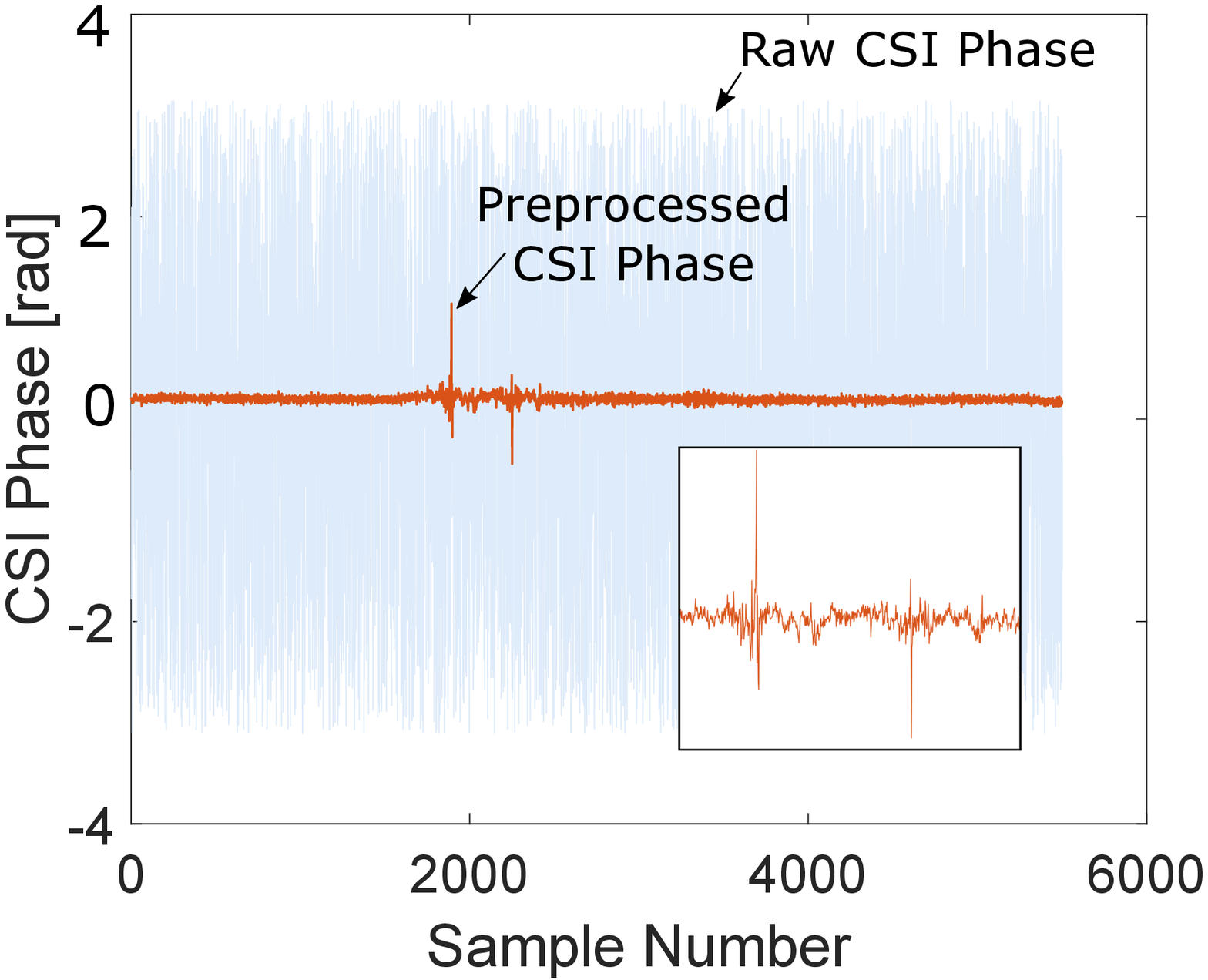}
\caption {Raw vs preprocessed CSI phase data for a passing vehicle.}
\label{fig:processed_phase}
\end{minipage}
\end{figure}


Figure~\ref{fig:pca_comp} shows the first PCA component compared with a filtered CSI stream. As shown, the PCA component more clearly represents the changes of CSI amplitude values induced by passing vehicles compared with the original CSI amplitude streams.

\subsubsection{Phase Preprocessing}
\label{sec:phase_preprocessing}

So far we described our approach to reducing the noise of CSI amplitude data. Now we present a method that mitigates the noise of CSI phase data.

We can express the measured CSI phase of subcarrier $c$ as the following~\cite{wu2015phaseu}.

\begin{equation}
\hat{\phi}_c = \phi_c - 2\pi\frac{k_c}{N_F}\alpha + \beta + Z.
\end{equation}

\noindent Here $\phi_c$ is the original phase; $k_c$ denotes the subcarrier index; $N_F$ is the Fast Fourier Transform size (64 for IEEE 802.11 a/g/n); and $Z$ is the measurement noise. Our objective is to remove $\alpha$ and $\beta$, which are the time lag and the phase offset at the receiver, respectively. We adopt a linear transformation to remove these noise factors~\cite{sen2012you}. Formally, we define the two variables $e_1$ and $e_2$ as follows.

\begin{equation}
e_1 = \frac{\hat{\phi}_F-\hat{\phi}_1}{2 \pi F}, \mbox{ } e_2 = \frac{1}{F}\sum_{1 \le c \le F}\hat{\phi}_c,
\end{equation}

\noindent Here $F$ refers to the last subcarrier index. Note that $F = 30$ because we use the Intel 5300 NIC which exports 30 subcarriers. We then use a linear transformation: $\hat{\phi}_f - e_1f - e_2$ to remove both the timing offset $\alpha$ and the phase offset $\beta$. We disregard the small measurement noise $Z$ in this calculation.


Figure~\ref{fig:processed_phase} shows both the raw and preprocessed CSI phase data (measured with sampling rates of 2,500 samples/sec). The result indicates that the proposed method successfully removes the random noise of CSI phase data and cleanly captures the original CSI phase data of a passing vehicle.

\subsection{Vehicle Detection}
\label{sec:vehicle_detection}

Given pre-processed CSI amplitude data $A$ and phase data $P$, we are ready to perform vehicle detection. In the vehicle detection module, $N_C$ passing vehicles are detected, and the corresponding CSI amplitude values $\matr{A}$ and CSI phase values $\matr{P}$ are captured from $A$ and $P$. More precisely, when a vehicle passes, distinctive peaks in CSI amplitude values are observed. To capture these peaks, a standard outlier detection based on the scaled median absolute deviation (MAD) technique is used. The scaled MAD is defined as follows.

\begin{equation}
\begin{aligned}
&\mbox{Scaled MAD}=c_{MAD} \cdot \mbox{median}(|a_i - \mbox{median}(A)|),\\ &\mbox{ } i=1,2,...,N.
\end{aligned}
\end{equation}

\noindent Here $c_{MAD} = \frac{-1}{\sqrt{2} \cdot \zeta(3/2)}$, where $\zeta$ is the inverse complementary error function. An $i$-th CSI amplitude value $a_i$ is considered as an outlier if it is more than three scaled MAD away from the mean.

\begin{algorithm}
  \caption{CSI Data Extraction Algorithm}
    \label{alg:algorithm1}
  \KwData{$A$ and $P$}
  \KwResult{$\matr{A}$ and $\matr{P}$}
  \Begin {
     $O$ $\longleftarrow outlier(A)$.\;
     \For{$i \leftarrow 1$ \KwTo $|O|$}{
        \uIf{$O[i]$ \&\& $!r$}{
           $s \longleftarrow i$, $f \longleftarrow i$, $r \longleftarrow$ TRUE.\;
        }
        \uElseIf {$O[i]$ \&\& $r$} {
           $f \longleftarrow i$.\;
        }
        \uElseIf {$!O[i]$ \&\& $r$} {
           \uIf {$f - i> \omega$} {
              $r \longleftarrow$ FALSE.\;
              \uIf {$(s - \delta_1) > 0$ \&\& $(f + \delta_2) < |O|$} {
                 $\matr{A} \longleftarrow A[(s-\delta_1)...(f+\delta_2)]$.\;
                 $\matr{P} \longleftarrow P[(s-\delta_1)...(f+\delta_2)]$.\;
              }
           }
        }
        \uElse {
           continue.\;
        }
     }
  }
\end{algorithm}

Once a sequence of outliers $O \subseteq A$ are detected, the vehicle detection module starts to extract $\matr{A}$ and $\matr{P}$ from $A$ and $P$, respectively. Let us explain first how to capture $\matr{A}$ from $A$. The starting $s$ and ending $f$ indices of the outliers are identified, and then CSI amplitude values in the range of [$a_{s-\delta_1}, a_{f+\delta_2}$] are extracted from $A = \{a_1, ..., a_N\}$, where $\delta_1$ and $\delta_2$ are system parameters. The former parameter is used to ensure that momentary CSI amplitude fluctuations are accounted for when a vehicle is very close to the line of sight (LoS) between TX and RX but not passed through it yet. Similarly, the latter is chosen to take into account the CSI amplitude changes made immediately after a vehicle passes the LoS. The CSI amplitude data for 783 vehicles were examined to determine $\delta_1$ and $\delta_2$. We found that $\delta_1 =500$ (0.25sec), and $\delta_2=500$ (0.25sec) were good enough to effectively capture $\matr{A}$ of all passing vehicles. Now we explain how to capture $\matr{P}$. Since the CSI phase data $P$ is synchronized with the CSI amplitude data $A$, the same technique is applied to capture $\matr{P}$, \emph{i.e.,} the CSI phase values in the range of [$p_{s-\delta_1}, p_{f+\delta_2}$] are extracted with the same parameters.

The pseudocode for extracting $\matr{A}$ and $\matr{P}$ is displayed in Algorithm~\ref{alg:algorithm1}. The function $outlier$ identifies outliers $O$ (Line 2). $O[i]$ is 1 if $a_i \in A$ is an outlier. $r$ is a flag used to indicate that the algorithm is in progress. We scan $O$ and find the starting index $s$ of the sequence of outliers (Lines 4-5). The algorithm then finds the ending index $f$ (Lines 6-7).  With the two indices, the length of the outlier sequence $f - i$ is compared to $\omega$ (Line 9). This is to ensure that the outlier sequence actually represents the CSI fluctuations for some period of time caused by a passing car, rather than just a momentary peak due to random noise. Finally, the amplitude values in the range of [$a_{s-\delta_1}, a_{f+\delta_2}$] and the phase values in the range of [$p_{s-\delta_1}, p_{f+\delta_2}$] are stored in $\matr{A}$ and $\matr{P}$, respectively (Lines 11-13). In our experiments, we used $\omega = 1,250$, \emph{i.e.,} 0.5 second).

\subsection{Vehicle Classification}
\label{sec:deep_learning}

We design a convolutional neural network (CNN) for vehicle classification. The basic idea is to correlate the time series of CSI amplitude and phase data by aggregating them as a single input image for CNN. More precisely, a $6 \times 2,500$ image is created in which the first three rows of the image represent the time series of extracted CSI amplitude values for three TX-RX antenna pairs (Note that there are 1 TX antenna, and 3 RX antennas in our experimental setting); The subsequent three rows of the image are the time series of extracted CSI phase values. These 6 CSI data sequences are exactly aligned in the image to allow for extraction of the hidden correlations between the CSI data sequences.

\begin{figure}[!htbp]
\centering
\includegraphics[width=.99\columnwidth]{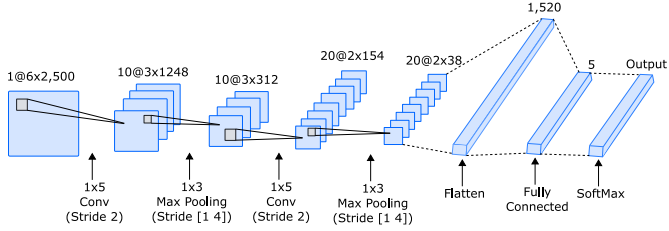}
\caption {Proposed CNN architecture.}
\label{fig:cnn}
\end{figure}

Figure~\ref{fig:cnn} shows the architecture of the proposed CNN. As shown, it consists of two layers of alternating Convolution (including the Batch Normalization and rectified linear unit (ReLU) sublayers), and Pooling such that the lower layer extracts basic features while the higher layer extracts more complex features~\cite{lecun2015deep}. In the following section, we describe the details of each layer.

\subsubsection{Convolutional Layer}
\label{sec:convolution_and_pooling}

The convolutional layer basically convolves input images by sliding the kernels vertically and horizontally. Denote the value at the $x$-th row and $y$-th column of the $j$-th input image in the $i$-th layer by $v_{i,j}^{x,y}$. To calculate $v_{i,j}^{x,y}$, the input images in the previous layer (\emph{i.e.,} ($i-1$)-th layer) are convolved with the kernels. The output of the convolutional layer is provided as input to an activation function $\sigma$ (\emph{i.e.,} the rectified linear unit (ReLU) function), and the result of the activation function forms the input image $v_{i,j}^{x,y}$ for the next layer. Mathematically, the convolution process is written as follows.

\begin{equation}
v_{i,j}^{x,y} = \sigma(\sum_m \sum_{k=0}^{\mathscr{L}_i-1}w_{i,j,m}^{k}v_{i-1,m}^{x,y+k} + g_{i,j}).
\end{equation}

\noindent Here $g_{i,j}$ is the bias for the $j$-th input image in the $i$-th layer; $w_{i,j,m}^{k}$ is the value of the kernel at the $k$-th position; $\mathscr{L}_i$ is the size of the kernel in the $i$-th layer; $m$ is an index that goes over the set of input images in the $(i-1)$-th layer.

Before providing the result of the convolutional layer as an input to the activation function, we pass the result through the normalization sublayer. The normalization sublayer is used to speed up the training process and reduce the sensitivity to the initial network configuration. Given a mini batch of the outputs of the convolution process $\mathfrak{B}=\{b_1, ..., b_N\}$, this sublayer performs normalization as the following:

\begin{equation}
\hat{b_i} = \frac{b_i - \mu_{\mathfrak{B}}}{\sqrt{\sigma_{\mathfrak{B}}^{2}+\epsilon}},
\end{equation}

\noindent where $\mu_{\mathfrak{B}} = \frac{1}{N}\sum_{i=1}^{N}b_i$, and $\sigma_{\mathfrak{B}}^2=\frac{1}{N}\sum_{i=1}^{N}(b_i - \mu_{\mathfrak{B}})^2$. Basically by subtracting the mean from the result of the convolutional layers and then dividing by the standard deviation, the normalization sublayer scales the result by a scale factor $\gamma_1$ and shifts by an offset $\gamma_2$. These two parameters $\gamma_1$ and $\gamma_2$ are learned during the training process.

After the convolution process followed by normalization, a nonlinear activation function $\sigma$ is executed, for which we adopt the rectified linear unit (ReLU) function. It basically performs a threshold operation to each element we obtain after the convolutional and batch normalization layer as follows: $\sigma = \mbox{ReLu}(v) = \mbox{max}(v,0)$.

\subsubsection{Max Pooling Layer}
\label{sec:max_pooling_layer}

In the max pooling layer, the resolution of the feature maps is decreased in order to prevent overfitting using the max pooling function defined as follows.

\begin{equation}
v_{i,j}^{x,y} = \mbox{max}_{1 \le k \le \mathscr{P}_i}(v_{(i-1),j}^{x,y+k}).
\end{equation}

\noindent Here $\mathscr{P}_i$ is the length of the pooling region in the $i$-th layer.

\subsubsection{Dropout and Fully Connected Layer}
\label{sec:dropout_layer}

While training the CNN model, we observed significant overfitting and decided to deploy the dropout layer to reduce the impact of overfitting. Basically, this layer randomly drops out an element of the results of the Max Pooling Layer with a fixed probability $p_{drop}$. In our experiments, we found that a drop out rate $p_{drop}$ of 0.6 gave the good results.

\subsubsection{Fully Connected Layer}
\label{sec:fully_connected_layer}

Followed by the two layers of alternating Convolution, Batch Normalization, ReLu, and Pooling sublayers is the Fully Connected Layer. This layer is basically the same as the regular neural network which maps the flattened feature into the output classes (\emph{i.e.,} five vehicle types) generating the scores for each output class. Finally, the output scores of the Fully Connected Layer is provided as input to the SoftMax layer in which the scores are converted into values in the range between 0 and 1 such that the sum is 1. This way the SoftMax layer represents the output as a true probability distribution.

\section{Experimental Results}
\label{sec:experimental_results}

\subsection{Experimental Setup}
\label{sec:experimental_setup}

\begin{wrapfigure}{r}{0.5\columnwidth}
\centering
\includegraphics[width=.5\columnwidth]{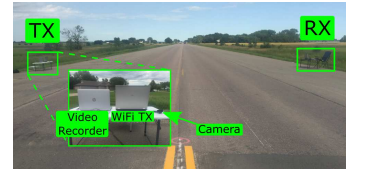}
\caption {Experimental setup.}
\label{fig:exp_setup}
\end{wrapfigure}


We deployed a prototype of $\mathsf{DeepWiTraffic}$ on a typical two-lane rural roadway (Figure~\ref{fig:exp_setup}). This rural roadway was selected to show that $\mathsf{DeepWiTraffic}$ serves effectively the objective of this paper to address the issue of deploying a large number of TMSs to cover huge miles of rural highways with reduced cost. $\mathsf{DeepWiTraffic}$ was implemented with two laptops (HP Elite 8730w). One was used as a WiFi transmitter; the other was used as a WiFi receiver. These laptops were equipped with 2.53GHz Intel Core Extreme CPU Q9300 processor, 4GB of RAM, and Intel 5300 NIC. $\mathsf{DeepWiTraffic}$ was executed on Ubuntu 14.04.04 (kernel version of 4.2.0-27). In addition to these two laptops, two more laptops were deployed to record the ground-truth video data--one recording incoming vehicles, and the other recording outgoing vehicles. The laptops for video recording were synchronized with the laptops for CSI data collection to facilitate the data labeling process.

\begin{table}[!htbp]
\center
\caption{Vehicle Types and Number of Samples}
\label{table:vehicle_number}
\begin{tabular}{ l|l|l|l }
\hlinewd{1.5pt}
\multicolumn{3}{ c| }{Vehicle Classification} & \# of Samples \\
\hline
\multirow{5}{*}{Car-like} & \multirow{2}{*}{Small} & Bike & 22 \\
 &  & Passenger Car & 238 \\ \cline{2-2}
 & \multirow{2}{*}{Medium} & SUV & 253 \\
 &  & Pickup Truck & 252 \\ \cline{1-2}
Truck-like & Large & Large Truck & 18 \\
\hlinewd{1.5pt}
\end{tabular}
\end{table}

WiFi CSI data and ground-truth video data were collected over a month period. The collected CSI data can be applied to general rural roadways as the environmental conditions of rural highways are usually very similar, \emph{i.e.,} two-lane with limited interference factors, \emph{etc.} This way, repeatedly collecting CSI data and building CNN models for similar rural roadway deployment can be avoided. To cope with rural roadways with significantly different environments such as the ones surrounded by dense trees, we can pre-build CNN models specifically for those environments and reuse the models. To further improve the efficacy of the proposed system, we can build separate CNN models for several typical weather conditions, \emph{e.g.,} rain and snow. We leave the task of building separate models for uncommon roadway environments and other weather conditions as future work.

We collected CSI data for 738 vehicles (Table~\ref{table:vehicle_number}). The collected CSI amplitude and phase data were labeled in accordance with the ground-truth video data. In labeling CSI data with different vehicle types, we referred to the FHA vehicle classification~\cite{FWHAClass}. Identifying a vehicle with more than two axles is known to be easy due to the distinctive vehicle body size. A major challenge lies in classifying vehicles with two axles since the body sizes of these vehicles are very similar. So we focus on classifying vehicles with two axles. Specifically according to the FHA vehicle classification~\cite{FWHAClass}, we used five vehicle classes: class 1 (moborcycle) class 2 (passenger car) class 3 (SUVs) class 4 (pickup truck), and class 5 (large truck). Here the large trucks mean a single unit with the axle count greater than or equal to three. As Table~\ref{table:vehicle_number} shows, we evaluated the performance of $\mathsf{DeepWiTraffic}$ for two other typical classification methods namely `car-like vs truck-like' classification~\cite{FWHAClass}, and `small, medium, large' classification~\cite{liang2015counting}.

\begin{table}[!htbp]
\center
\caption{Hyper Parameters Used for CNN}
\label{table:parameters}
\begin{tabular}{ l|l}
\hlinewd{1.5pt}
Parameter Type & Value \\ \hline
Solver & \pbox{20cm}{Stochastic Gradient Descent \\ with Momentum (SGDM) Optimizer} \\
Dropout Rate & 60\% \\
Shuffle Frequency & Every Epoch \\
Validation Data & 30\% \\
Input Image Size & 6 $\times$ WINDOW\_SIZE \\
WINDOW\_SIZE & 2,500 \\
L2 Regularization & None \\
\hlinewd{1.5pt}
\end{tabular}
\end{table}

Table~\ref{table:parameters} summarizes the hyper parameters ued for training the CNN model. As shown, we used 70\% of the collected CSI data to train the CNN model, and the rest for testing. We compared the performance of $\mathsf{DeepWiTraffic}$ with that of support vector machine (SVM) and k-nearest neighbor (kNN). In particular we used the following five features in training the SVM and kNN models:  (1) the normalized standard deviation
(STD) of CSI, (2) the offset of signal strength, (3) the period of the vehicle motion, (4) the median absolute deviation (MAD), (5) interquartile range (IR) according to ~\cite{wang2017wifall} which exploited WiFi CSI for fall detection.

Most TMSs are evaluated with two metrics, vehicle detection accuracy and vehicle classification accuracy. The detection accuracy is defined as $\frac{\mbox{Number of detected vehicles}}{\mbox{Number of passing vehicles}}$. The classification accuracy is defined as $\frac{\mbox{Number of correctly classified vehicles}}{\mbox{Number of detected vehicles}}$. The performance of $\mathsf{DeepWiTraffic}$ is evaluated based on these metrics.

\subsection{Detection Accuracy}
\label{sec:detection_accuracy}

In this experiment, 778 vehicles were detected out of 783 passing vehicles with a total of 24 false positives, achieving the detection accuracy of 99.4\%. This result coincides with the literature that most recent TMSs have high vehicle detection accuracy. In particular, the near-100\% detection accuracy of $\mathsf{DeepWiTraffic}$ can be attributed to the PCA component that makes the induced CSI amplitude values of a passing vehicle even more distinctive (Figure~\ref{fig:pca_comp}).

\subsection{Classification Accuracy}
\label{sec:classification_accuracy}

\begin{table*}[!htbp]
\center
\caption{Classification Accuracy}
\label{table:accuracy}
\begin{tabular}{ l|l|l||l|l|l|l|l|l|l|l|l}
\hlinewd{1.5pt}
\multicolumn{3}{ c|| }{Classification} & \multicolumn{3}{ c| }{SVM} & \multicolumn{3}{ c| }{kNN} & \multicolumn{3}{ c }{$\mathsf{DeepWiTraffic}$} \\ \hline \hline

\multirow{4}{*}{Car-like} & \multirow{2}{*}{Small} & Bike & \multirow{4}{*}{99.3\%} & \multirow{2}{*}{85.7\%} & 81.3\% & \multirow{4}{*}{99.2\%} & \multirow{2}{*}{46.5\%} & 77.2\% & \multirow{4}{*}{100.0\%} & \multirow{2}{*}{91.1\%} & 97.2\% \\
 &  & Passenger Car & & & 75.9\% & & & 54.5\% & & & 91.1\%\\ \cline{2-2} \cline{5-5} \cline{8-8} \cline{11-11}

 & \multirow{2}{*}{Medium} & SUV & & \multirow{2}{*}{85.8\%} & 50.6\% & & \multirow{2}{*}{92.8\%} & 47.8\% & & \multirow{2}{*}{94.1\%} & 83.8\%\\
 &  & Pickup Truck & & & 75.5\% & & & 42.5\% & & & 83.3\%\\ \cline{1-2} \cline{4-5} \cline{7-8} \cline{10-11}

Truck-like & Large & Large Truck & 98.0\% & 96.2\% & 95.5\% & 92.8\% & 91.1\% & 90.4\% & 100.0\% & 100.0\% & 99.7\%\\ \hline \hline

\multicolumn{3}{ c|| }{Average} & \textbf{98.7\%} & \textbf{89.2\%} & \textbf{75.8\%} & \textbf{96.0\%} & \textbf{76.8\%} & \textbf{62.5\%} & \textbf{100.0\%} & \textbf{95.1\%} & \textbf{91.1\%}\\

\hlinewd{1.5pt}
\end{tabular}
\end{table*}

The classification accuracy of $\mathsf{DeepWiTraffic}$ was measured and compared with that of SVM and kNN-based approaches. In this experiment, we randomly selected 30\% of the passing vehicles as the validation set for SVM, kNN, and $\mathsf{DeepWiTraffic}$. This experiment was repeated 1,000 times and the average classification accuracy was calculated.

The results are summarized in Table~\ref{table:accuracy}. All classifiers did a good job in classifying vehicles into two classes of the `car-like' and `truck-like' classes. The classification accuracy, however, was degraded as we increased the number of classes from 2 to 3, and to 5. The reason is that it is more difficult to classify correctly vehicles with similar body sizes. A notable observation was that the classification accuracy of SVM and kNN dropped very sharply with the increased number of classes (98.7\% $\rightarrow$ 75.8\% for SVM, and 96.0\% $\rightarrow$ 62.5\% for kNN), while $\mathsf{DeepWiTraffic}$ maintained relatively high classification accuracy (100\% $\rightarrow$ 91.1\%). The high classification accuracy of $\mathsf{DeepWiTraffic}$ can be attributed to the automatic extraction of useful features rather than relying on fixed parametrization of input data. Despite this strength of the modern machine learning technique, we observed that distinguishing vehicles with very similar body sizes, \emph{i.e.,} SUVs and pickup trucks, is still challenging, resulting in the classification accuracy of only 83.8\% for SUVs and 83.8\% for pickup trucks. Yet, considering the high classification accuracy over all vehicle types, $\mathsf{DeepWiTraffic}$ shows very promising performance comparable to recent non-intrusive solutions.

\begin{figure}[!htbp]
\centering
\includegraphics[width=.9\columnwidth]{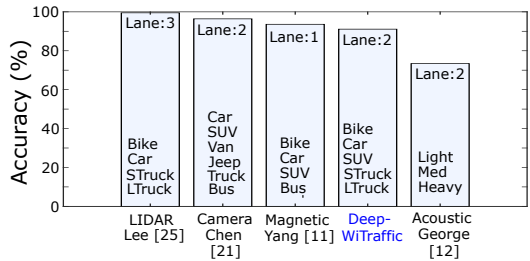}
\caption {Comparison with other non-intrusive solutions.}
\label{fig:comparison}
\end{figure}

Figure~\ref{fig:comparison} shows the classification accuracy of other solutions in comparison with $\mathsf{DeepWiTraffic}$. The figure displays the average classification accuracy achieved with each solution type. It also shows vehicle types for classification and the number of lanes. The LIDAR and camera-based systems achieve very high classification accuracy. However, due to the high cost, these solutions are more appropriate for urban highways with many lanes. The result also shows that compared with other non-intrusive solutions such as magnetic-based and acoustic-based TMSs, $\mathsf{DeepWiTraffic}$ achieves high classification accuracy. It is worth to note that the result is significant in that the classification accuracy of $\mathsf{DeepWiTraffic}$ is attained with a larger number of vehicle types and same or greater number of lanes, compared with the magnetic-based and acoustic-based solutions.



\subsection{Classification Accuracy Per Lane}
\label{sec:effect_of_lane}

\begin{table*}[!htbp]
\center
\caption{Classification Accuracy Per Lane}
\label{table:accuracy_per_lane}
\begin{tabular}{ l|l|l||l|l|l|l|l|l|l|l|l}
\hlinewd{1.5pt}
\multicolumn{3}{ c|| }{Classification} & \multicolumn{3}{ c| }{Lane 1} & \multicolumn{3}{ c| }{Lane 2} & \multicolumn{3}{ c }{Combined Lane} \\ \hline \hline

\multirow{4}{*}{Car-like} & \multirow{2}{*}{Small} & Bike & \multirow{4}{*}{100.0\%} & \multirow{2}{*}{91.1\%} & 97.2\% & \multirow{4}{*}{100.0\%} & \multirow{2}{*}{90.3\%} & 97.0\% & \multirow{4}{*}{99.8\%} & \multirow{2}{*}{79.6\%} & 95.5\% \\
 &  & Passenger Car & & & 91.1\% & & & 87.5\% & & & 81.6\%\\ \cline{2-2} \cline{5-5} \cline{8-8} \cline{11-11}

 & \multirow{2}{*}{Medium} & SUV & & \multirow{2}{*}{94.1\%} & 83.8\% & & \multirow{2}{*}{93.7\%} & 83.1\% & & \multirow{2}{*}{80.0\%} & 76.3\%\\
 &  & Pickup Truck & & & 83.3\% & & & 80.0\% & & & 66.5\%\\ \cline{1-2} \cline{4-5} \cline{7-8} \cline{10-11}

Truck-like & Large & Large Truck & 100.0\% & 100.0\% & 99.7\% & 100.0\% & 100.0\% & 99.1\% & 99.5\% & 99.0\% & 92.1\%\\ \hline \hline

\multicolumn{3}{ c|| }{Average} & \textbf{100.0\%} & \textbf{95.1\%} & \textbf{91.1\%} & \textbf{100.0\%} & \textbf{94.7\%} & \textbf{89.3\%} & \textbf{99.7\%} & \textbf{86.2\%} & \textbf{82.4\%} \\

\hlinewd{1.5pt}
\end{tabular}
\end{table*}

Another interesting research question that we answer here is: how does the lane (more specifically, the distance between a passing car and the WiFi receiver) affect the performance of $\mathsf{DeepWiTraffic}$. To answer this question, we created CNN models separately for each lane simply based on the lane tag that we put during the labeling process; Note that this does not require additional CSI data collection for different lanes. The results for different CNN models for Lane 1, Lane 2, and Combined Lanes are summarized in Table~\ref{table:accuracy_per_lane}. We found that the effect of lanes was negligible when vehicles were classified into the `car-like' and `truck-like' classes. This results show that if vehicles for classification have significantly different shapes the distance between the car and the WiFi receiver has marginal effect.

In contrast, the classification accuracy for the combined lanes degraded by 8.9\% and 8.7\% for the `S,M,L' classes and the individual vehicle classes, respectively. A possible explanation is that CSI data for each lane has distinctive characteristics, and when the CSI data for the two lanes are combined, these unique features are obscured, resulting in degraded classification accuracy. An easy solution to address this discrepancy is to create a CNN model individually for each lane, perform classification for each CNN model, and use the classification result with a higher probability.

Another interesting observation is that the classification accuracy for Lane 1 is slightly higher than that for Lane 2. The reason is, as illustrated in Figure~\ref{fig:system_overview}, when a passing vehicle is close to the WiFi receiver, WiFi signals are spaced relatively more widely. As a result, the physical characteristics of the vehicle body are more effectively captured leading to higher classification accuracy.

\section{Related Work}
\label{sec:related_work}

\begin{table*}[t]
\center
\caption{Vehicle Classification Techniques}
\label{table:classification_vehicle}
\begin{tabular}{ l|l|l|l|l|l }
\hlinewd{1.5pt}
Classification & Approach & Publication & Vehicle Class & Cost & Accuracy\\ \hline
Intrusive & Piezoelectric sensor & Rajab~\cite{rajab2014vehicle} & \pbox{6cm}{motorcycles, passenger vehicles, two axle for tire unit, buses, two axles six tire single units, three axles single units, four or more axles single unit, four axles single trailer, five axles single trailer, seven axles single trailer, seven or more axles multi-trailer} & medium & 86.9\% \\ \cline{2-6}
& Magnetometer & Bottero~\cite{bottero2013wireless} & \pbox{6cm}{car, van, truck} & medium & 88.0\% \\ \cline{3-6}
& & Xu~\cite{xu2018vehicle} & \pbox{6cm}{hatchbacks, sedans, buses, and multi-purpose vehicles} & medium & 95.4\% \\ \cline{2-6}
& Loop Detector & Meta~\cite{meta2010vehicle} & \pbox{6cm}{car, van, truck, bus, motorcycle} & high & 94.2\% \\ \cline{3-6}
& & Jeng~\cite{jeng2014high} & \pbox{6cm}{motorcycles, passenger cars, other two-axle four-tire single unit vehicles, buses, two-axle, six-tire single-unit trucks, tree-axle single-unit trucks, four or more axle single-unit trucks, four or fewer axle single-trailer trucks, five-axle single-trailer trucks, six or more axle single-trailer trucks, five or fewer axle multi-trailer trucks, six-axle multi-trailer trucks, seven or more axle multi-trailer trucks} & high & 92.4\% \\ \cline{1-6}
Non-intrusive & Camera & Chen~\cite{chen2012vehicle} & \pbox{6cm}{car, van, bus, motorcycle} & medium/high & 94.6\% \\  \cline{3-6}
& & Bautista~\cite{bautista2016convolutional} & \pbox{6cm}{jeep, sedan, truck, bus, SUV, and van} & medium/high & 96.4\% \\ \cline{2-6}
& Infrared + ultrasonic sensors & Odat~\cite{odat2017vehicle} & \pbox{6cm}{sedan, pickup truck, SUV, bus, two wheeler} & low/medium & 99.0\%  \\ \cline{2-6}
& Magnetic Sensors & Wang~\cite{wang2014easisee} & \pbox{6cm}{bicycle, car, minibus} & low/medium & 93.0\% \\ \cline{3-6}
& & Yang~\cite{yang2015vehicle} & \pbox{6cm}{motorcycle, passenger car (two-box and saloon), SUV, bus} & low/medium & 93.6\%  \\ \cline{2-6}
& Acoustic Sensors & George~\cite{george2013vehicle} & \pbox{6cm}{heavy (truck bus), medium (car, jeep, van), light (auto rickshaws, two wheelers)} & low/medium & 73.4\% \\ \cline{2-6}
& LIDAR & Lee~\cite{lee2012side}\cite{lee2015using} & \pbox{6cm}{motorcycle, passenger vehicle, passenger vehicle pulling a trailer, single-unit truck, single-unit truck pulling a trailer, and multi-unit truck} & high & 99.5\%  \\ \cline{2-6}
& RF & Al-Husseiny~\cite{al2012rf} & \pbox{6cm}{human, car} & low & 89.0\% \\ \cline{3-6}
& & Kassem~\cite{kassem2012rf} & \pbox{6cm}{empty street, stationary car,
and moving car} & low & 100.0\%  \\ \cline{3-6}
& & Haferkamp~\cite{haferkamp2017radio} & \pbox{6cm}{car, truck} & low & 99.0\%  \\ \cline{1-6}
Off-roadway & UAV & Liu~\cite{liu2015fast} & \pbox{6cm}{car, truck} & medium & 98.2\% \\ \cline{3-6}
& & Tang~\cite{tang2017arbitrary} & \pbox{6cm}{seven vehicle types, such as car, truck, bus, etc. (specific type not specified)} & medium & 78.7\% \\ \cline{2-6}
& Satellites & Audebert~\cite{audebert2017segment} & \pbox{6cm}{pick up, van, truck, car} & high & 80.0\% \\
\hlinewd{1.5pt}
\end{tabular}
\end{table*}

Vehicle detection and classification is a key functionality of TMS~\cite{meta2010vehicle}. While the vehicle detection accuracy is known to be very high, the vehicle classification accuracy differs substantially depending on techniques applied. As such, this section is focused on presenting a comprehensive review on vehicle classification techniques. Vehicle classification methods are divided largely into three categories: intrusive, non-intrusive, and off-roadway approaches. Table~\ref{table:classification_vehicle} summarizes the properties of existing vehicle classification schemes including sensor types, vehicle types, classification accuracy, and the cost.


In intrusive solutions, sensors (\emph{e.g.,} piezoelectric sensors~\cite{rajab2014vehicle}, magnetometers~\cite{bottero2013wireless}\cite{xu2018vehicle}, vibration sensors~\cite{stocker2014situational}, loop detectors~\cite{meta2010vehicle}) are installed on or under a roadway. As it is shown in Table I, intrusive approaches are capable of classifying a large selection of vehicle types with high classification accuracy leveraging close contact with passing vehicles that allow for securing high-precision sensor data. The major problem of these solutions, however, is the high cost for installation and operation. Especially when sensors are installed under the pavement, the cost increases prohibitively. The maintenance cost is also non-negligible as it incurs extra cost for constructor safety assurance.

Due to the high cost of intrusive solutions, non-intrusive approaches have received attention. In non-intrusive solutions, sensors are deployed on a roadside or over the road obviating the construction and maintenance cost. A most widely adopted sensor for non-intrusive solutions is a camera~\cite{chen2012vehicle}\cite{bautista2016convolutional}. Significant advances in imaging technologies and image processing techniques based on machine learning algorithms gave a birth to precise camera-based TMSs~\cite{niessner2017investigations}. As Table I shows, the classification accuracy of camera-based TMSs is very high. However, achieving high classification accuracy is still challenging at night, under severe weather conditions, and when there are obstacles that obstruct the clear view. There are other sensors such as magnetometers~\cite{wang2014easisee}\cite{yang2015vehicle}, accelerometers~\cite{ma2014wireless}, and acoustic sensors~\cite{george2013vehicle} that have been used in non-intrusive TMSs. Table I shows that state-of-the-art systems based on these sensors have good classification accuracy. However, the low-fidelity information that these sensors provide requires coordinated and precise positioning of multiple of those sensors. As such, minor errors in positioning or adjusting sensing directions may result in poor classification performance. To address the drawbacks, more advanced sensors such as LIDAR (Laser Infrared Detection And Ranging)~\cite{lee2012side}\cite{lee2015using} were considered. While these advanced technologies allow for very high classification accuracy, the cost is very high.

Off-roadway solutions utilize cameras mounted on UAVs~\cite{tang2017arbitrary} or satellites~\cite{audebert2017segment} for vehicle classification. As shown in Table I, the classification accuracy of off-roadway approaches is not very high. The low classification accuracy of off-roadway solutions is mainly attributed to the small image size. However, off-roadway approaches are appropriate when the user needs to cover a large area.

Recently solutions exploiting wireless signals have received much attention. RF-based solutions utilize RF signals for traffic detection and identification~\cite{kassem2012rf}\cite{al2012rf}\cite{haferkamp2017radio}. General RF signals, however, convey only limited information, most prominently the signal strength. Consequently, the RF-based solutions are limited to simple vehicle classification (\emph{e.g.,} car or truck) or just vehicle detection. On the other hand, WiFi CSI allows for more sophisticated analysis of the channel condition (\emph{e.g.,} combined effect of scattering, fading, and power decay) via concurrent transmissions of multiple subcarriers for each pair of TX-RX antennas. In comparison with our previous work~\cite{won2017witraffic}, we achieve significantly higher vehicle classification accuracy for diverse types of vehicles by developing advanced CSI data processing techniques and by integrating with state-of-the-art machine learning algorithms.

\section{Conclusion}
\label{sec:conclusion}

We presented $\mathsf{DeepWiTraffic}$, a WiFi-based TMS that addresses the critical cost problem for covering the huge rural road network. Unique channel properties induced by passing vehicles are utilized to achieve high vehicle detection and classification accuracy. Extensive experiments were conducted based on a huge amount of WiFi CSI data and ground-truth video data were collected. The results show that $\mathsf{DeepWiTraffic}$ achieved near 100\% vehicle detection accuracy, and average classification accuracy of 91.1\% in comparison with state-of-the-art non-intrusive solutions. Our future work includes more extensive CSI data collection from different traffic environments, performance evaluation under more complex settings such as higher vehicle density, people moving around, and varying weather conditions.

\section*{Acknowledgments}

This work was supported by the DGIST R\&D Program of the Ministry of Science and ICT (19-EE-01). This research was also supported in part by the Competitive Research Grant Program (CRGP) of South Dakota Board of Regents (SDBoR).

\bibliographystyle{IEEEtran}
\bibliography{mybibfile}



\end{document}